\documentclass[twocolumn,footinbib,a4paper]{revtex4}
\usepackage{amssymb}
\usepackage{amsmath}
\usepackage{graphicx}
\newcommand{\be}{\begin{eqnarray}}
\newcommand{\ee}{\end{eqnarray}}

\begin{document}

\title{Electronic properties, doping and defects in chlorinated silicon nanocrystals}

\author{A.~Carvalho}
\email{aicarvalho@ua.pt}
\affiliation{Department of Physics, I3N, University of Aveiro, Campus Santiago, 3810-193 Aveiro, Portugal}

\author{S.~\"Oberg}
\affiliation{Department of Engineering Sciences and Mathematics, Lule{\aa} University of Technology, Lule{\aa}~S-97187, Sweden}

\author{M.~J.~Rayson}
\affiliation{Department of Engineering Sciences and Mathematics, Lule{\aa} University of Technology, Lule{\aa}~S-97187, Sweden}

\author{P. R. Briddon}
\affiliation{Electrical, Electronic and Computer Engineering, University of Newcastle upon Tyne, Newcastle upon Tyne NE1 7RU, United Kingdom}

\begin{abstract}
Silicon nanocrystals with diameters between 1 and 3~nm and surfaces passivated by chlorine
or a mixture of chlorine and hydrogen were modeled using density functional theory, and
their properties compared with those of fully hydrogenated nanocrystals.
It is found that fully and partially chlorinated nanocrystals are stable,
and have higher electron affinity, higher ionization energy and lower optical absorption energy threshold.
As the hydrogenated silicon nanocrystals, chlorinated silicon nanocrystals 
doped with phosphorus or boron require a high activation energy to transfer an electron or hole, respectively,
to undoped silicon nanocrystals. The electronic levels of surface dangling bonds are similar
for both types of surface passivation, although in the chlorinated silicon nanocrystals some fall off the narrower gap.
\end{abstract}

\maketitle

\section{Introduction}

The manipulation of the silicon surface and its ability to interact with molecules and radicals
 is gaining importance in view of the use of silicon nanostructures in
 hybrid inorganic-organic colloids and other functional materials.
With a large surface-to-volume ratio, free-standing silicon nanocrystals (NCs) 
are ideal to explore the surface functionality.
They can be obtained by ultrasonic dispersion of porous silicon,\cite{heinrich-S-1992}
liquid phase synthesis by reduction of SiCl$_4$,\cite{bley-JACS-118-12461}
or plasma processes.\cite{kortshagen-in-pavesi-book,mangolini-AM-19-2513,oda-ACIS-71-31,oda-JJAP-36-4031,nozaki-N-18-235603,gresback-N-22-305605,li-L-19-8490}
Nonthermal plasma synthesis is an efficient method for production of particles
of mono-disperse sizes and lists,
amongst other advantages, suppressed particle coagulation  and
selective heating of particles through energetic surface reactions.\cite{kortshagen-in-pavesi-book}
Additionally, it offers the possibility of dopant (P,B) incorporation during growth.\cite{nakamine-JAP-50-025002,stegner-PRB-80-165326,lechner-JAP-104-053701,pi-APL-92-123102}

Although silane is usually chosen as a precursor for
plasma synthesis, SiCl$_4$ has also been suggested as a cheaper and safer alternative.\cite{nozaki-N-18-235603,gresback-N-22-305605}
Nanocrystals grown from a SiCl$_4$/H$_2$/Ar mixture are terminated 
with a mixture of chlorine and hydrogen,
with variable fractions depending on the plasma composition and reactor pressure.\cite{nozaki-N-18-235603,gresback-N-22-305605}

The fraction of surface Cl on silicon nanocrystals grown by this or other methods
can also be increased by Cl$_2$ plasma etching, treatment with a solution of PCl$_5$ on
chlorobenzene or with di-, tri-, and tetrachlorosilane gases,
procedures already in use for industrial processing
of silicon single crystal substrates.\cite{buriak-CR-102-1271}
This leads to the formation of mono-, di-  and trichloride terminations at low temperatures
($<400^\circ$C).\cite{schnell-PRB-32-8057}
 Monochloride is the most stable, remaining at higher temperature.
The adsorption of chlorine and SiCl$_n$ on flat silicon surfaces
has been extensively studied by theoretical methods.\cite{wijs-PRL-78-4877,sakurai-SS-493-143,sakurai-JCG-237-21,humbird-JAP-96-971,rivillon-JVSTA-23-1100,solares-JACS-128-3850,mohapatra-PRB-38-12556}
The adsorption energy of Cl$_2$ on a reconstructed Si(100) surface was found to be­ 5.4 eV,\cite{wijs-PRL-78-4877}
giving a Si-Cl bond energy of about 4~eV. 
On Si(111) surfaces, the Si-Cl bond energy is similar,
the calculated values ranging between 3.5 and 4.2 eV.\cite{sakurai-SS-493-143}
In both cases, the barrier for chlorine diffusion is about 1~eV,
and desorption takes place in the SiCl$_2$ form.\cite{wijs-PRL-78-4877,sakurai-SS-493-143}
Further, it was found that with increasing chlorine supply the structure of the chlorinated Si(111) surface 
suddenly changes, with a transition from a monochloride phase to a polychloride phase.\cite{sakurai-SS-493-143}
The adsorption of chlorine induces Cl-related Cl-Si bonding states below the top of the valence band.\cite{schnell-PRB-32-8057}

The reactive Si-Cl surface bonds are convenient for surface functionalization
with alkene and amine groups.\cite{veinot-in-pavesi-book,soria-L-27-2613}
Although Si-Cl bonds are stronger than Si-H bonds,
the Cl atoms, with a higher affinity for electrons, 
can more easily receive an electron from the highest occupied
molecular orbital (HOMO) of the other reactant during the interaction.
This additional electron is partially localized on the shallowest $p$ orbital of the Cl
radical, resulting in  Cl$^-$ being released.
Forming Cl$^-$ in the transition state is energetically more favorable
than breaking Si-H bonds, leading to lower activation energy barriers 
for grafting in Si-Cl bonds, even for partial Cl coverages.\cite{soria-L-27-2613}

Additionally, the presence of chlorine changes the optical and electronic properties
of the material, opening exciting possibilities for surface-driven electronic structure
engineering. 
Previous electronic structure calculations have found that chlorine-covered nanocrystals
have a lower gap between occupied and unoccupied electron energy levels 
and higher electron affinity than hydrogen-covered nanocrystals.\cite{martinez,ma-JPC-115-12822} 
Thus, it is possible that partial or full surface chlorination can be
used to control the position of the electronic levels for specific applications.

Given the interest on Cl-terminated nanocrystals, 
both for subsequent surface conversion or for electronic structure engineering,
theoretical information on the stability, electronic and optical properties of silicon
nanocrystals is of great interest.
Therefore, we have carried out a detailed theoretical study 
to compare the properties of Cl-terminated silicon nanocrystals
with 1-3~nm of diameter with H-terminated nanocrystals in the same size range.
The first-principles methodology will be described in Section~I.
The structure and energetics of perfect nanocrystals with Cl-, H- and mixed terminations
will be considered in Section~II, 
and their electronic and optical properties will be given in Section III.
Section IV is dedicated to doped and defective Cl- and H- terminated clusters.
Finally, Section V discusses the relevance of the results.

\section{Methodology \label{sec:method}}

The electronic structure of the nanocrystals was analyzed using first-principles calculations
based on density functional theory,
with a pseudopotential approach, as implemented in the {\sc aimpro} code.~\cite{briddon-PSSB-217-131,rayson-CPC-178-128} 
The local density approximation\cite{briddon-PSSB-217-131} was used for the exchange and correlation energy.
Core electrons were accounted for by using the pseudopotentials of Hartwigsen {\it et al.}.\cite{hartwigsen-PRB-58-3641}

Kohn-Sham orbitals were expanded on a localized basis set consisting of atom-centered Cartesian Gaussian orbitals
with angular momentum up to $l=2$, 
as described in Ref.~\cite{goss-TAP-104-69}.
For the core silicon atoms, we used a contracted basis set with 13 functions per atom (44G*),
including a polarization function with $l=2$, optimized for
bulk silicon. A basis of the same size, optimized for SiH$_4$, was used for hydrogen.
For chlorine, an uncontracted basis set with four $l=0$ and twelve $l=1$ functions per atom was used.
Convergence tests for silane, tetrachlorosilane and Si$_{87}$H$_{76}$/Si$_{87}$Cl$_{76}$ nanocrystals 
show that these bases offer an excellent compromise between accuracy and computational effort,
specially for large nanocrystal diameters, where the electronic structure becomes 
increasingly bulk-like.
For the worst case, the SiH$_4$ and SiCl$4$ molecules,
Si-H and Si-Cl bond lengths (Table~\ref{tab:SiX4}) are converged respectively within 0.006 and 0.015~\AA,
bond energies are converged within 0.1 and 0.5 eV, respectively,
and the Kohn-Sham gaps are converged with 0.22 and 0.09~eV, respectively.
They are also in good agreement with previous LDA calculations.

Total energy calculations were performed using finite real space boundary conditions.
The optical absorption cross section was calculated using periodic boundary conditions, ensuring a minimum distance of at least 10~{\AA} between replicas of the system.
In this case, the charge density was expanded in a plane wave basis set with an energy cutoff of 350~Ry.

The equilibrium geometry of the nanocrystals was found by a relaxation
of all the atomic coordinates using a conjugate gradient algorithm.


The optical absorption cross-section was calculated in the long-wavelength dipole approximation
using the Kohn-Sham eigenvalues $E^n$ and eigenvectors $|\psi^k\rangle$.
The matrix elements of ${\bf r}$ are evaluated using the momentum operator plus the commutator of the
non-local part of the pseudopotential.\cite{read-PRB-44-13071}
The Brillouin Zone sampling was restricted to the $\Gamma$ point.
The electronic temperature used as parameter in the Fermi-Dirac distribution was 0.1~eV/$k_B$, 
where $k_B$ is the Boltzmann constant, and the Gaussian broadening used was 0.05~eV.

\begin{table}
\caption{
Bond length, bond enthalpy,
Kohn-Sham gap ($\Delta E_{\rm KS}=E_{\rm LUKS}-E_{\rm HOKS}$)
and vertical excitation energy
of the SiH$_4$ and SiCl$_4$ molecules.
\label{tab:SiX4}} 
\begin{ruledtabular}
\begin{tabular}{lllllll}

\multicolumn{2}{l}{property} &$l_{{\rm Si}-X}$ (\AA)      &$\Delta H_b$ (eV)    & $\Delta E_{\rm KS}$ &$\Delta E^*$& \\
\hline
SiH$_4$    & This work       &1.49                       &3.8                   &7.93                    & \\
		& Prev. calc.     &1.50\footnotemark[1]       & 3.5\footnotemark[2]  & 7.93 \footnotemark[5]& 8.76\footnotemark[5],9.26\footnotemark[6]\\
           &Exp.             &1.48\footnotemark[8]       & 3.3,\footnotemark[3] 3.2\footnotemark[4]      & \\[0.5em]
SiCl$_4$   & This work       &2.03                       & 4.3                     &6.37                 &6.67\\
           & Prev. calc.     &                           &                          &                    &9.14\footnotemark[7]\\
           &Exp.             &2.02\footnotemark[8]       & 4.0\footnotemark[4]      &                    &\\
\end{tabular}
\end{ruledtabular}

\footnotetext[1]{All-electron LDA calculation from Ref.\cite{jones-PRB-64-125203}}  
\footnotetext[2]{LDA calculation from Ref.\cite{erzenhof-JCP-110-5029} }  
\footnotetext[3]{From Ref.\cite{feller-JPCA-103-6413}}  
\footnotetext[4]{From the heats of formation in Ref.\cite{JANAF}}  
\footnotetext[5]{GGA-PBE calculation from Ref.\cite{degoli-PRB-69-155411}}  
\footnotetext[6]{B3LYP calculation from Ref.\cite{lehtonen-PRB-72-085424}}  
\footnotetext[7]{Discrete variational X$\alpha$ calculation\cite{isikawa-JCP-94-6740}}  
\footnotemark[8]{From Ref.~\cite{NISTdata}}

\end{table}

\section{Structure and energetics}

\begin{table}
\caption{Atomic composition, diameter and symmetry of the Si$_nX_m$ nanocrystals studied, 
where $X\in$\{H,Cl\}.\label{tab:clusters}}
\begin{ruledtabular}
\begin{tabular}{llll}
$n$ & $m$  & $d$ (nm) & sym.\\ 
\hline
	35	&36      &1.1 & T$_d$\\
	87	&76      &1.5 & T$_d$\\
	244	&144     &2.1 & T$_d$\\
	275	&172     &2.2 & T$_d$\\
	286	&170     &2.2 & D$_{3d}$\\
	377	&196     &2.4 & T$_d$ \\
	513	&252     &2.7 & T$_d$\\
	717	&300     &3.0 & T$_d$\\
\end{tabular}
\end{ruledtabular}
\end{table}

\subsection{Structure}
The nanocrystals used in this study were obtained by cutting
an approximately spherical core out of a 
perfect silicon crystal and passivating the surface dangling bonds
with Cl or H atoms. 
The cutoff diameter can be estimated as $d=[3n/(4\pi)]^{1/3}a_0$,
where $n$ is the number of silicon atoms
 and $a_0$ is the calculated lattice parameter of bulk silicon (5.39~\AA).
The surface silicon atoms were four-fold coordinated
and had mono- or di-hydride/chloride termination.
The number of Si and Cl or H atoms in each nanocrystal is given in
Table~\ref{tab:clusters}.
All the nanocrystals were centered at an atomic site, with exception
of the Si$_{286}X_{170}$ nanocrystals, which were centered at a bond-center.

\begin{figure*}
\includegraphics{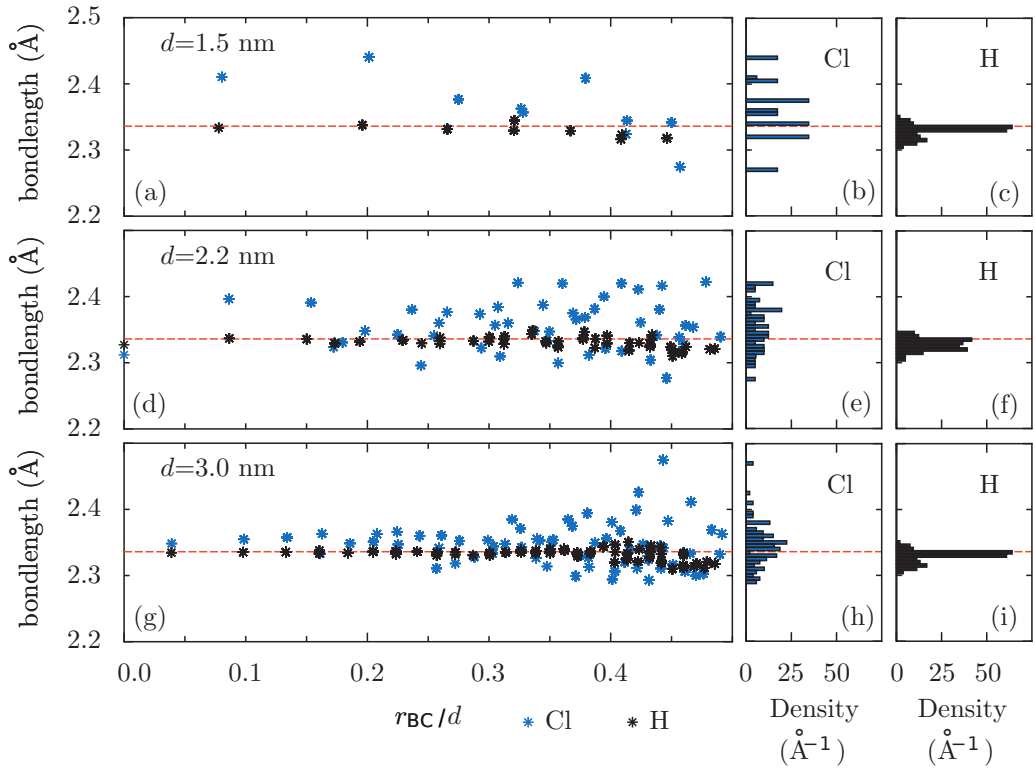}
\caption{ Calculated Si-Si bondlengths as a function of the distance
of the bondcenter to the center of the nanocrystal (${r}_{\rm BC}$),
and respective bond length histograms, for three nanocrystal
diameters: (a-c) 1.5 nm, (d-e) 2.2 nm (D$_{3d}$) and (f-h) 3.0 nm.
The dashed line represents the bulk Si bondlength.\label{fig:l}}
\end{figure*}

After atomic relaxation, all silicon atoms remain four-fold coordinated,
and the lengths and angles of the Si-Si bonds are close to those of the bulk crystal,
specially those at largest distances from the surface [Fig.~\ref{fig:l}-a),d) and g)].
Both Cl- and H-terminated nanocrystals maintain a marked crystalline character,
characterized by a discrete radial pair distribution function relative to the 
nanocrystal center.
However, there are quantitative differences between Cl- and H-covered nanocrystals.
As highlighted in Fig.~\ref{fig:l}, the bondlength distribution is much broader
for the chlorinated nanocrystals. 
For example, for $d=1.5$~nm, the average Si-Si bondlength of the Cl- and H-terminated nanocrystals
deviates only $-$0.01 and +0.02\AA, respectively, from the calculated
bulk Si-Si bondlength (2.34~\AA),
but the standard deviation is four times larger for the latter (Table~\ref{tab:blt-x}).
Curiously, the bond angles are closer to the bulk value for the chlorinated nanocrystal,
since the effective radius of the Cl atom is closer to that of the silicon atom.

The Si-H surface bonds are elongated about 0.2 \AA with respect to their length in the SiH$_4$ molecule
(1.49 \AA, to be compared with the experimental value 1.48 \AA\cite{NISTdata}).
However, the Si-Cl bondlengths are very close to those of SiCl$_4$ 
(2.03~\AA, to be compared with the experimental value of 2.02~\AA\cite{NISTdata}).

We now analyze the structure of nanocrystals with a mixed Cl/H surface.
The fraction $x$ of Cl atoms was varied between 0 and 1 ($x\in${0.25, 0.50, 0.75}), 
where $x$ is the ratio between 
the number of Cl atoms and the total number of Cl and H atoms.
For each $x$, the bondlengths and angles were averaged over 24 
randomly generated samples (Table~\ref{tab:blt-x}).
We notice that for the three intermediate $x$ fractions
the distribution of Si-Si bondlengths is more narrow than for $x=1$ (Cl-covered nanocrystal),
whereas the bond angle distribution is more narrow than for $x=0$ (H-covered nanocrystal).

\begin{table}
\caption{Average Si-Si, Si-Cl and Si-H bond lengths ($\bar{l}_{X-Y}$) 
and angles ($\bar{\alpha}$),
and their standard deviations, for a nanocrystal with diameter $d=1.5$~nm,
as a function of the surface Cl coverage ratio ($x$).
Lengths are given in Angstrom, and angles in degrees.\label{tab:blt-x}
The experimental bulk Si-Si bondlength is 2.35~\AA.}
\begin{ruledtabular}
\begin{tabular}{lcccccc}
$x$                    &   0    & 0.25  & 0.5    &   0.75   &  1     & bulk Si\\      
\hline 
$\bar{l_{\rm Si-Si}}$  & 2.33   & 2.33  & 2.33   &  2.34     & 2.36  & 2.34\\ 
$\bar{l_{\rm Si-Cl}}$  &  --    & 2.10  & 2.09   &  2.09     & 2.07  & \\
$\bar{l_{\rm Si-H}}$   & 1.72   & 1.69  & 1.68   &  1.69     &  --   & \\
$\Delta l_{\rm Si-Si}$ & 0.0084 & 0.0095& 0.014  &  0.024    & 0.04  & \\
$\Delta l_{\rm Si-Cl}$ &  --    & 0.012 & 0.044  &  0.072    & 0.0098& \\
$\Delta l_{\rm Si-H}$  & 0.39   & 0.38  & 0.36   &  0.38     & --    & \\ 
$\bar{\alpha}$         & 103.2  & 106.0 & 108.2  &  109.3    & 109.4 & 109.5\\
$\Delta\alpha$         & 20.4   & 15.9  & 11.2   &  7.5      &  4.0  &   \\
\end{tabular}
\end{ruledtabular}
\end{table}

\begin{figure}
\includegraphics{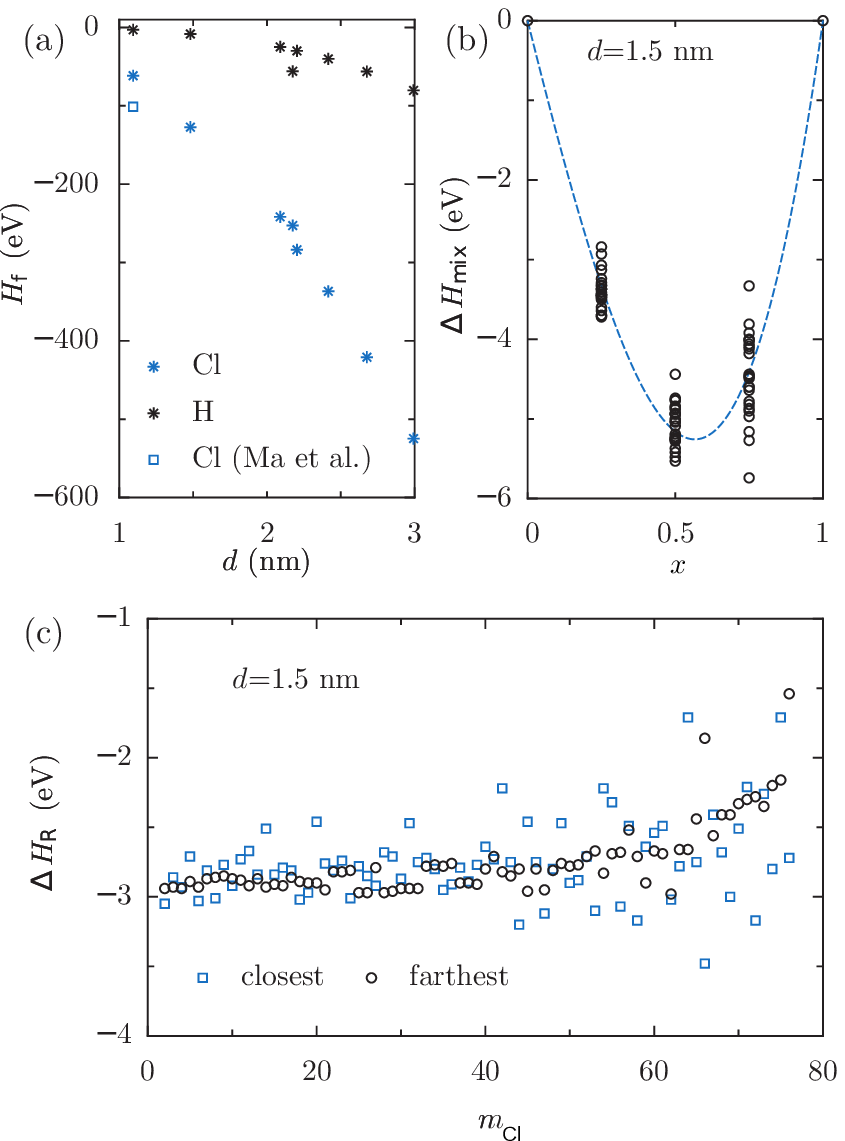}
\caption{Stability of Si$_n$Cl$_{m_{\rm Cl}}$H$_{m_{\rm H}}$ nanocrystals.
(a) Enthalpy of formation of Cl- and H- terminated nanocrystals as a function of the diameter,
(b) Mixing enthalpy (Eq.~\ref{eq:mix}) of $d=1.5$~nm nanocrystals with mixed surface as 
a function of the Cl fraction $x=m_{\rm Cl}/(m_{\rm Cl}+m_{\rm H})$,
and (c) comparison of the Cl replacement enthalpy (Eqs.~\ref{eq:R1}--~\ref{eq:R2}),
for an even distribution of Cl (placement of each Cl in the farthest position 
from the existing $m_{\rm Cl}-1$ Cl atoms)
and for a concentrated distribution of Cl 
(placement of Cl in the nearest position to the existing $m_{\rm Cl}-1$ Cl atoms),
for $d=1.5$~nm.\label{fig:H}}
\end{figure}

\subsection{Formation and reaction enthalpies}

It is important to know how the passivation with Cl affects the stability of the nanocrystals.
Although highly metastable structures can be prepared out of equilibrium,
for example in non-thermal plasmas, the enthalpy of formation is an useful to characterize
the stability in equilibrium and estimate reaction energies.
It is therefore useful to evaluate the enthalpy of formation of the nanocrystals with Cl-covered
surface or with mixed Cl and H surface passivation.

We calculated the formation enthalpies at $T=0$, defined as the enthalpy change
relative to the standard phases of Si (crystalline Si), Cl (molecular Cl$_2$) and H (molecular H$_2$)
\begin{equation}
H_f=E_{\rm NC}(n,m_{\rm Cl},m_{\rm H})-nE_{\rm Si}-\frac 1 2m_{\rm Cl}E({\rm Cl_2})-\frac 1 2 m_{\rm H}E({\rm H_2})
\end{equation}
where $E_{\rm NC}(n,m_{\rm Cl},m_{\rm H})$ is the calculated total energy of 
Si$_n$Cl$_{m_{\rm Cl}}$H$_{m_{\rm H}}$,
and $E({\rm Si})$, $E(\rm Cl_2)$ and $E(\rm H_2)$ are the total energy per atom 
of crystalline silicon and the total energies of the Cl$_2$ and H$_2$ molecules, respectively,
calculated using the same approximations. 
When the formation reaction is exothermic, $H_f$ is negative.

We note that the total energies of small molecules, in particular H$_2$,
are not accurately calculated using the LDA approximation. 
Thus, our calculated formation enthalpies of SiCl$_4$ and SiH$_4$ are underestimated:
we obtain $-$7.1 and $-$0.2~eV, respectively, whereas the experimental values
are $-6.6$ and $-0.4$~eV\cite{JANAF}.
However, this error often cancels out when calculating reaction energies.
For example, the enthalpy change for the hydrogen replacement reaction
\begin{equation}\rm SiH_4+Cl_2\rightarrow SiH_3Cl+HCl,\end{equation}
which is $-$2.83~eV in our calculation, is only underestimated by $0.06$~eV 
(relative to the value obtained from the experimental heats of formation\cite{JANAF}).
Thus, the calculated formation enthalpies can still be used to draw qualitative conclusions.

Nanocrystal formation enthalpies are shown in Fig.~\ref{fig:H}-a).
Cl-covered clusters have lower $H_f$ than H-covered clusters, typically by 1.1-1.6~eV per surface Si-H or Si-Cl bond.
This difference is larger than the difference between the bond energies in the SiCl$_4$ and SiH$_4$ molecules (Table~\ref{tab:SiX4}).
It is also larger than the errors in $H_f(\rm SiCl_4)$ and $H_f(\rm SiH_4)$.

The enthalpy of formation of the clusters with mixed Cl/H surface 
follows very closely a linear interpolation of the endpoints $x=0$ and $x=1$.
In analogy with the alloys, we can define a mixing enthalpy characterizing the deviation from linearity:
\be
\Delta H_{\rm mix}=E_{\rm NC}(n,m_{\rm Cl},m_{\rm H})-\left[E_{\rm NC}(n,m_{\rm Cl}+m_{\rm H},0)\right.\\ \nonumber
\left.-E_{\rm NC}(n,0,m_{\rm Cl}+m_{\rm H}).\right]\label{eq:mix}
\ee
This is given in Fig~\ref{fig:H}-b) for $d=1.5$~nm. 
The mixing enthalpy $\Delta H_{\rm mix}$ is negative and smaller by one or two orders of magnitude
than the enthalpy of formation.
If, near the temperature at which the Cl- and H- atoms become mobile,
the mixing free energy remains negative, 
this means that a binary system with Si$_n$Cl$_m$ and Si$_n$H$_m$ moieties
will be unstable against the formation of a mixed Si$_n$Cl$_{m_{\rm Cl}}$H$_{m_{\rm H}}$ ensemble.

Since the effective radius of the Cl atoms is much larger than that of H atoms,
an additional question is whether steric effects prevent Cl atoms from occupying neighboring positions,
even preventing complete chlorination altogether.
To investigate this, we calculated the enthalpy change associated with the hydrogen replacement reactions,
\be
{\rm Si}_n{\rm Cl}_{m_{\rm Cl}-1}{\rm H}_{m_{\rm H}+1}+{\rm Cl}_2\rightarrow
{\rm Si}_n{\rm Cl}_{m_{\rm Cl}}{\rm H}_{m_{\rm H}}+{\rm HCl}\label{eq:R1}
\ee
for $m_{\rm Cl}<n$, which is given by
\be
\Delta H_{\rm R}=E_{\rm NC}(n,m_{\rm Cl},m_{\rm H})+ E({\rm HCl})-\\ \nonumber
\left[E_{\rm NC}(n,m_{\rm Cl}-1,m_{\rm H}+1)-E({\rm Cl_2})\right]\label{eq:R2}
\ee
for $d=1.5$~nm nanocrystals.
Two opposite situations were considered.
Starting with Si$_{87}$ClH$_{75}$, we first created an even distribution of Cl 
by placing each additional Cl atom in one of the surface sites 
(position {\bf r}) minimizing the objective function
\begin{equation}
f({\bf r})=\sum_{i=1}^{m_{\rm Cl}-1}|{\bf r}-{\bf r}_i|^{-1}.
\end{equation}
This results in a sequence of clusters where Cl replacing for H
takes place at the position further away from all the other Cl atoms.
The enthalpy changes for this sequence of replacements
are compared with those obtained for a concentrated Cl distribution,
where each Cl atom is placed as close as possible to the atoms of the same species
(thus maximizing $f({\bf r})$).
The results [Fig~\ref{fig:H}-c)]
show that there is no clear energetic preference for the first process,
although the distribution of the enthalpies of replacement is smoother and narrower.
Moreover, the enthalpies of replacement stay approximately constant up to 50\% coverage,
showing only a slight increase for higher $x$.
So, there is in principle no reason why complete Cl coverage would not be attainable.

\section{Electronic and optical properties}
\subsection{Analysis of the Kohn-Sham states}

Let us start by analyzing the electronic structure of the SiCl$_4$ and SiH$_4$ molecules,
as represented by the Kohn-Sham eigenstates and eigenvalues.
Although these quantities have only an auxiliary role in DFT,
their analysis is useful to understand the bonding and
the contribution of Cl and H atoms to the ground state ans excited states.
The highest occupied state of the SiCl$_4$ molecule
is the $2t_1$ state, followed closely by the 2$e$ and 8$t_2$ states.
The HOKS is completely localized on the Cl atoms (formed by Cl $3p$ states).
The 2$e$ and 8$t_2$ states also have a localization of less than 10\% on the Si atom.
In contrast, the HOKS of SiH$_4$, which is the 2$t_2$ state, is a bonding state 41\% localized in the Si atom.
The lowest unoccupied state of the SiCl$_4$ molecule is the 8$a_1$ state, 
followed 1.7~eV above by the 9$t_2$ state,
whereas the LUKS of SiH$_4$ is 3$t_2$. Both are partially localized on Si:
54\% in the case of the SiCl$_4$ LUKS, 65 \% in the case of the SiH$_4$ LUKS.

\begin{figure}
\includegraphics[width=8.5cm]{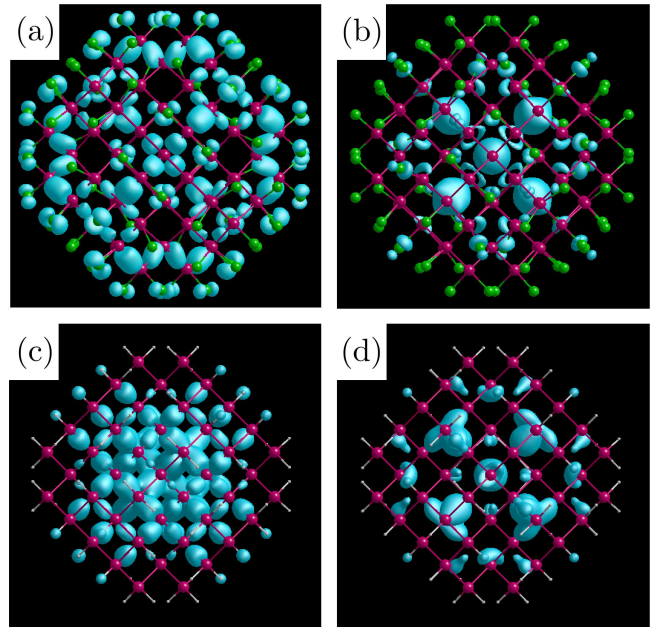}
\caption{Charge density associated with the HOKS and LUKS levels:
(a) HOKS of Cl-covered nanocrystal
(b) LUKS  of Cl-covered nanocrystal
(c) HOKS of H-covered nanocrystal
(d) LUKS of H-covered nanocrystal.
The isosurface value is $3\times10^{-3}$ and $8\times10^{-4}$
for HOKS and LUKS states, respectively.
\label{fig:KS}}
\end{figure}

As the number of Si atoms increases, the highest occupied Kohn-Sham (HOKS) state and the LUKS
start to resemble the bulk silicon valence and conduction band states,
but in the case of the Cl-covered clusters the Cl $3p$ character is maintained.
Figure~\ref{fig:KS} depicts the charge density associated with the 
HOKS (triplet) and LUKS states for $d=1.5$~nm.
The localization of those gap-edge states on
the surface atoms is greater for the Cl-covered
cluster, specially for the HOKS state 
(the fractional HOKS localizations on Cl/H are respectively 40 and 9\% for the Cl- and H-covered NCs).
The contribution of the Cl 3$p$ atomic orbitals to the HOKS state
is evident in the shape of the charge density isosurface near the surface of the nanocrystal,
which resembles the SiCl$_4$ 2$t_1$ state.
Similarly, near the surface the LUKS state bears some resemblance to 
the SiCl$_4$ LUKS ($8a_1$) state.

The calculated HOKS-LUKS gap ($E_{\rm KS}$) of the SiH$_4$ and SiCl$_4$ molecules
is respectively 7.93 and 6.96~eV (Table~\ref{tab:SiX4}).
Although these are not far from the experimental absorption energy thresholds,
which are respectively 8.99~eV\cite{suto-JCP-84-1160} and 8.84~eV,\cite{ibuki-JCP-106-4853} 
there are several reasons why they cannot be compared directly to experiment.
Firstly, the HOKS and LUKS states of SiCl$_4$ are 2$t_1$ and 8$a_1$, respectively,
and the HOKS$\rightarrow$LUKS optical transition is forbidden by symmetry;
the lowest allowed transition, $2t_1\rightarrow9t_2$,
corresponds to an eigenvalue energy difference of 8.1~eV.
Moreover, 
the Kohn-Sham states change considerably in the excited state,
and so do Coulomb, exchange and correlation interactions.
Moreover, the threshold energy of the absorption spectra of both molecules
is a Rydberg transition ($4s\rightarrow8t_2$)\cite{suto-JCP-84-1160,ibuki-JCP-106-4853}.
These excitonic effects leading to Rydberg states are not described by the 
ground state DFT.
To our knowledge, Rydberg transitions have not been resolved for undoped silicon nanocrystals.

\begin{figure}
\includegraphics[width=8.5cm]{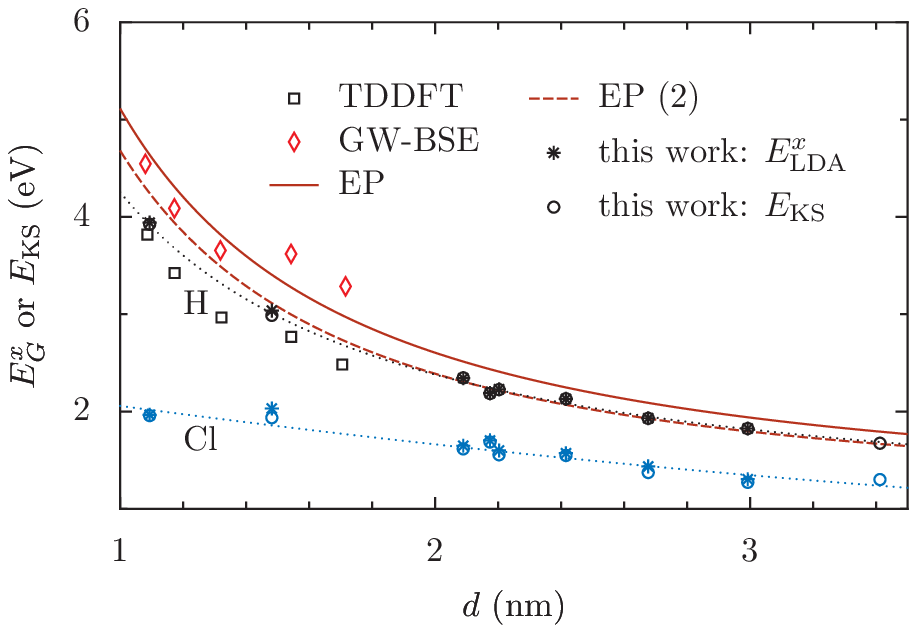}
\caption{Excitation energy (*) and Kohn-Sham bandgap ($\circ$) 
as a function of the nanocrystal diameter obtained from the present work. 
Data from previous calculations are shown for comparison: 
lowest excitation energy excitations calculated by solving the Bethe-Salpeter 
equation (GW-BSE) from Ref.\onlinecite{tiago-PRB-73-205334},
or obtained by time-dependent LDA (TDLDA), also from Ref.\onlinecite{tiago-PRB-73-205334},
calculated using empirical pseudopotentials without inclusion of the
Coulomb interaction between electron and hole (EP) or
including this contribution (EP2), from Ref.\onlinecite{proot-PRL-61-1948}.
Dotted lines are a guide to the eye.
\label{fig:Ex}}
\end{figure}

For silicon nanocrystals with diameters between 1 and 3~nm, 
the lowest excitation energy, obtained by calculating the difference
between the total energies of each nanocrystal in the ground state
and in the first excited state (at the ground state geometry),
$E_{\rm LDA}^x=E_G^1-E_G^0$,
differs little from $E_{\rm KS}$ (Fig.~\ref{fig:Ex}).
This means that, upon excitation, the change in the electrostatic 
interaction energy 
(resulting from the interaction between electron, hole, and image charges)
is partially canceled by the change in the exchange and correlation energy.

The excitation energy ($E_x$) of hydrogenated nanocrystals has been previously
calculated at different levels of theory. 
As illustrated in Fig.~\ref{fig:Ex},
our results are in good agreement with previous calculations with empirical potentials,
and differ less than 1~eV from GW gaps. 
The reason why the LDA HOKS-LUKS bandgap is a good approximation for the
excitation energy is clarified by Delerue {\it et al.},
who have proved that the differences between the corrections to the
self-energy in bulk and in the nanocrystal ($\delta\Sigma E$), 
are nearly canceled out by the Coulomb interaction between electron and hole ($E_{C}$).
As a result,
$E_x\simeq E_{\rm KS}+\delta\Sigma b$, where $\delta\Sigma b$ is
the bulk self-energy correction, which is about 0.6-0.7~eV for
the LDA approximation. 
In this work, we will assume that this correction is independent
of the nanocrystal surface, thus justifying the comparison between
Cl- and H-covered nanocrystals using 
the values directly obtained from first-principles.

\begin{figure}
\includegraphics[width=8.5cm]{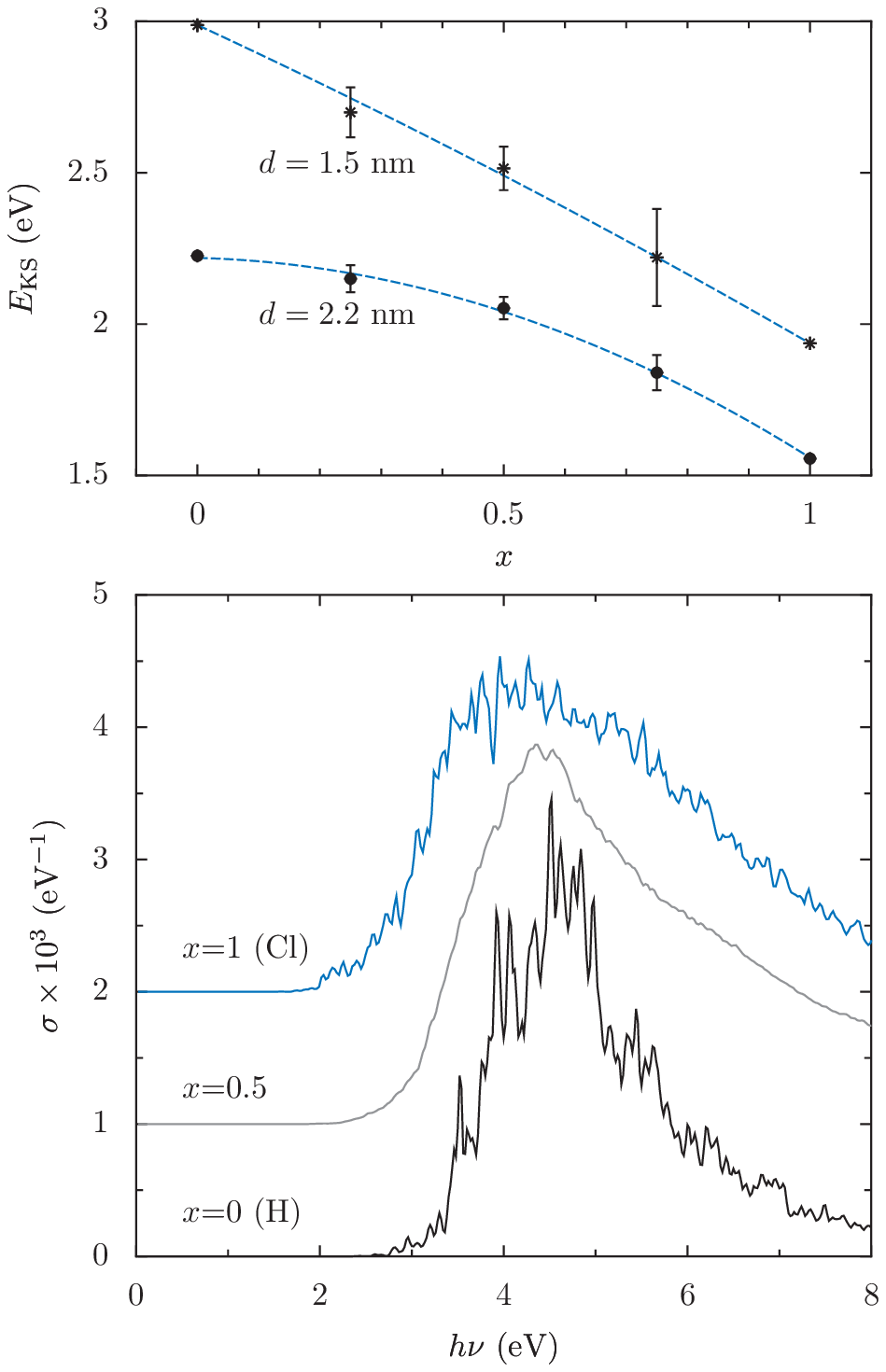}
\caption{Kohn-Sham bandgap (a) and calculated absorption spectra (b)
of $d=2.2$~nm silicon nanocrystals for different
fractions of Cl coverage ($x=m_{\rm Cl}/(m_{\rm Cl}+m_{\rm H})$).
The eigenvalue gaps for fractional $x$ were obtained by averaging over 24 samples,
and the errorbars represent the standard deviation of the results. 
The absorption spectra for $x=0.5$ was obtained by averaging over 10 samples.
\label{fig:opt}}
\end{figure}

The minimum excitation energy is lower for the 
Cl-covered nanocrystal than for the H-covered nanocrystal.
This follows the lower effective confinement volume for the HOKS and LUKS
states in the Cl-covered nanocrystals. 
The difference is greater for the smaller diameters,
amounting to about 1~eV for $d\sim1.5$~nm.
With increasing $d$, the gap of the Cl-covered clusters decreases
almost linearly in this size range, 
whereas that of the H-covered nanocrystals it varies approximately with $d^{-1.2}$.
The average gap of nanocrystals with mixed Cl- and H-coverage varies monotonically
between those of the Cl- and H-covered clusters of the same size  (Fig.~\ref{fig:opt}).
The variation in the gap distribution for each set of samples with the same $d$ and $x$
is not negligible, and is represented in Fig.~\ref{fig:opt} by the errorbars.

\subsection{Optical spectra}

We have calculated the optical absorption cross section 
directly using the Kohn-Sham eigenstates
and eigenvalues, as described in Sec.~\ref{sec:method} 
Previous theoretical work has shown that optical spectra can, 
in a good approximation, 
be obtained from the Kohn-Sham eigenvalues, since
only minor charge rearrangements occur between
ground and the low excited states.\cite{delley-PRB-47-1397}
As expected, the absorption threshold energy is lower for the Cl-covered
cluster (Fig.~\ref{fig:opt}). 
However, the energies close to $E_{\rm KS}$ have very small or vanishing
oscillator strengths. 
The threshold is steeper for the H-covered nanocrystal and
for the nanocrystal with mixed surface, were the symmetry is broken,
than for the Cl-covered nanocrystals.
Overall, the absorption band in the 2-6~eV range is broader
for the Cl-covered nanocrystal.

\subsection{Ionization energy and electron affinity}

\begin{figure}
\includegraphics{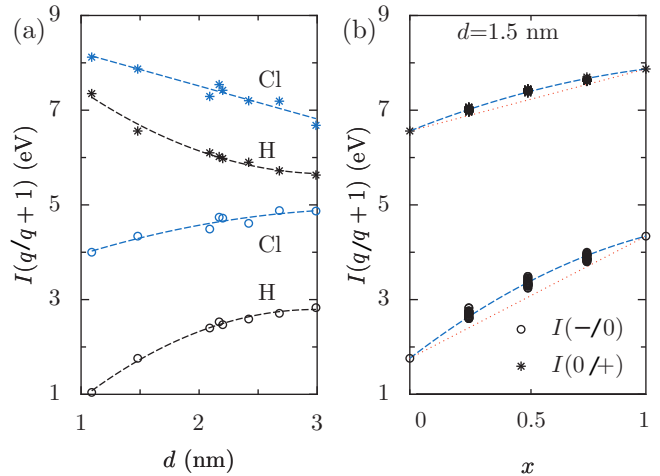}
\caption{Ionization energy [$I(0/+)$] and electron affinity [$I(-/0)$],
(a) for Cl- and H-covered nanocrystals, as a function of the diameter, and
(b) for $d=1.5$~nm nanocrystals with mixed surface 
( Si$_{87}$Cl$_{m_{\rm Cl}}$H$_{m_{\rm H}}$), as a function of
the Cl fraction $x=m_{\rm Cl}/(m_{\rm Cl}+m_{\rm H})$.
The absorption spectra for $x=0.5$ was obtained from the average
of ten random surface configurations.
\label{fig:I}}
\end{figure}

The ionization energy was obtained from the total energies of
neutral and charged clusters,
\be I(q/q+1)=E(q+1)-E(q), \ee
where $q$ is the charge. 
The results are shown in Fig.~\ref{fig:I}.
The electron affinity of the chlorinated nanocrystals is higher by 2-3~eV,
for the whole range of $d$ considered, reflecting the higher affinity
for electrons of Cl. The ionization energy is also higher, but only
by about 1~eV.

\section{Defects}

\subsection{Dopants}

\begin{figure}
\includegraphics[width=8.5cm]{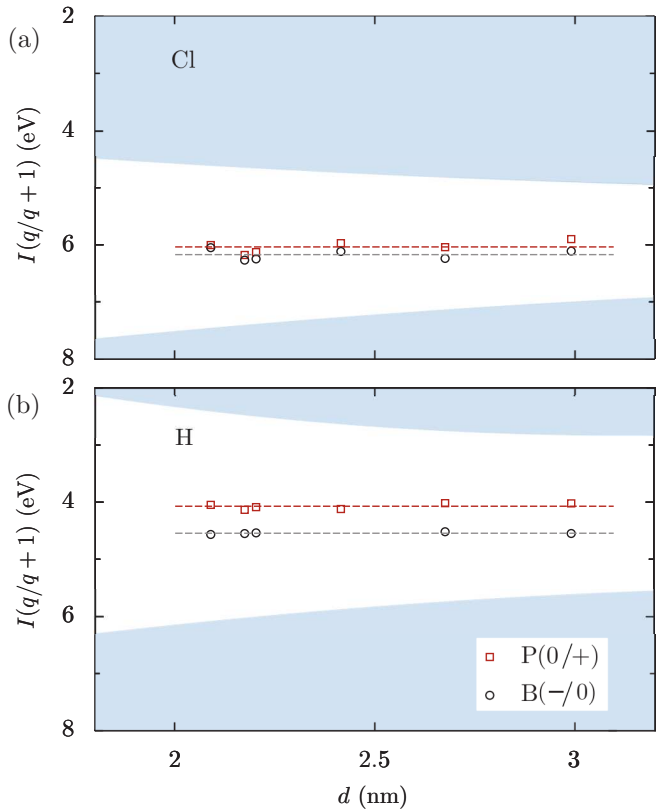}
\caption{
Ionization energy [$I(0/+)$] and electron affinity [$I(-/0)$]
of doped and undoped (a) Cl-covered nanocrystals and 
(b) H-covered nanocrystals, as a function of the diameter.
Shaded areas represent energies lower than the electron affinity
or higher than the ionization energy of the undoped nanocrystal.
The axis were inverted for easier visualization.
\label{fig:BP}}
\end{figure}

Further, we compare the ionization energy/electron affinity
of doped nanocrystals with that of the pristine (undoped) nanocrystals.
This comparison is relevant when doping nanocrystal composites 
where only a small fraction of nanocrystals encloses one or more dopant atoms.
In that case, the ideal is that a nanocrystal doped with a shallow donor
(for example P) has ionization energy $I(0/+)$ very close to the electron affinity
$I(-/0)$ of the undoped nanocrystal.
Ideally, $I_{\rm P}(0/+)-I_{\rm UD}(-/0)$, where the subscripts label
the doped and undoped nanocrystals 
should be comparable to $kT$, where $T$ is the temperature and $k$ Boltzmann constant.
The reverse is true for shallow acceptors.
However, this does not happen either for Cl- or H-covered nanocrystals
with $d$ between 2 and 3~nm (Fig.~\ref{fig:BP}).
This is due to the carrier confinement and appearance of image charges,
which were already extensively discussed for H-covered nanocrystals.\cite{lanoo-PRL-74-3415,zhou-PRB-71-245308,melnikov-PRL-92-046802,chan-NL-8-596,cantele-PRB-72-113303}

\begin{figure}
\includegraphics[width=8.5cm]{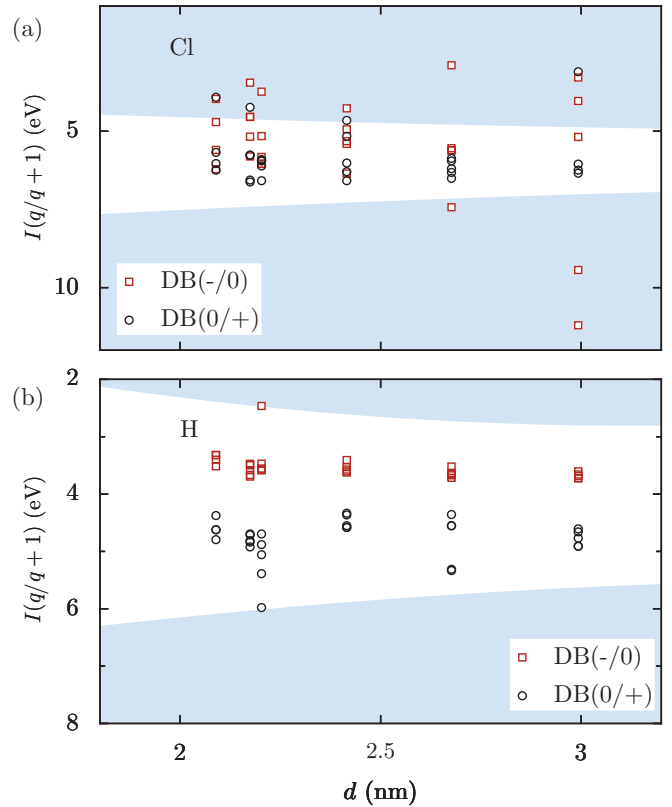}
\caption{Incomplete
Ionization energy [$I(0/+)$] and electron affinity [$I(-/0)$]
of surface dangling bonds on (a) Cl-covered nanocrystals and 
(b) H-covered nanocrystals, as a function of the diameter.
Dangling bonds were formed on all symmetry-nonequivalent 
di-hydride terminations.
Shaded areas represent energies lower than the electron affinity
or higher than the ionization energy of the undoped nanocrystal.
The axis were inverted for easier visualization.
\label{fig:DB}}
\end{figure}

\subsection{Dangling bonds}

The position of the $I(q/q+1)$ levels of surface dangling bonds (DBs)
relative to the gap edge states of the pristine nanocrystals has also been 
compared for both types of surface.
We considered only dangling bonds on di-chloride or di-hydride surface silicon
atoms i.e., those that in the pristine nanocrystal were attached to two surface
terminators.
In $d=1.5$~nm nanocrystals, monohydride DBs, although in average higher in energy by 0.9~eV,
have similar properties.
For each NC size, 
there are several nonequivalent surface Si atoms with di-chloride (or di-hydride) 
terminations where dangling bonds can form. 
The geometry and energy of each of the respective defects was optimized independently.

First, we note that the $(-/0)$ and $(0/+)$ level positions
do not display a clear trend with the nanocrystal size.
The DB levels are very dependent on the particular defect geometry, 
specially on the Cl-covered nanocrystals (Fig.~\ref{fig:DB}). 
The main difference between Cl- and H-covered nanocrystals is that, 
as the gap of the former is smaller, 
some of the DB levels fall outside the gap. 
That does not happen for the H-covered nanocrystals.

\section{Discussion}

Silicon NCs with chlorinated or mixed surface have two main potential uses: 
as an intermediate material for further surface functionalization and modification,
and as an electronic or optoelectronic material. 
The energies and electronic properties obtained in this study may be useful to design or tune both types of applications.

We have show that the formation enthalpy of the chlorinated Si-NCs is more negative than that of 
the hydrogenated Si-NCs, relative to the standard states of Cl and H. 
However, this does not mean that the former will in general be more stable against reaction.
In fact, chlorinated Si-NCs have very high electron affinity, and can easily trap electrons
to the LUMO state, which is partially localized on the surface Cl orbitals (Fig.~\ref{fig:KS}).
This leaves Cl more susceptible to removal and substitution by a foreign radical.
It is interesting to note that although mixed Cl and H surfaces have a negative mixing enthalpy,
there is no strong repulsion between nearest-neighbor Cl atoms, specially for small coverage ratios. 
Thus, if the nanocrystals are kept neutral, there is in principle the possibility of engineering
next-neighbor surface replacements using hydrogenated NCs with selected Cl substitutions.

Chlorine can be used to modify intentionally the electronic and optical properties of the Si-NCs.
Chlorinated Si-NCs also have a smaller gap between occupied and unoccupied electron levels.
As a result, the threshold energy for optical absorption is redshifted.
The absorption edge can be varied by changing the Cl coverage ratio.

As both the electron affinity and ionization energy are greater than those of hydrogenated
silicon nanocrystals, the Cl surface coverage ratio can be used to tune the alignment between
the Si-NC states and the bands of other materials in heterojunctions.
Chlorinated Si-NCs doped with P [or B] have ionization energy [electron affinity] levels 
quite distant from the ($-$/0) [(0/+)] levels of pristine nanocrystals with the same size.
This also happens with hydrogenated Si-NCs, making it difficult for a doped NC to donate free carriers
to undoped nanocrystals.
However, as the gap of the chlorinated Si-NCs 
shifted to lower energies, P(0/+) and B($-$/0) levels
are also shifted relative to the vacuum level,
in comparison with the hydrogenated crystals.
This knowledge may be useful to design heterojunctions with doped Cl-terminated Si-NCs as one of the components.
As in the hydrogenated Si-NCs, dangling bonds will act as exciton recombination centers.

\acknowledgements
The computations were performed on resources
provided by the Swedish National Infrastructure for Computing (SNIC)
at KTH (Lindgren), Ume\aa\ University (Akka), University of Aveiro (Blafis) and University of Coimbra (Milipeia).
This work was supported by 
the Calouste Gulbenkian Foundation, 
the Marie Curie Program REG/REA.P1(2010)D/22847 (SiNanoTune) and 
FCT Portugal (SFRH/BPD/66258/2009 and PTDC/FIS/112885/2009).

\end{document}